\input{aipcheck}

\documentclass[
    ,final            
  ]
  {aipproc}

\layoutstyle{8x11double}

\begin{document}

\title{Analysing solar-like oscillations with an automatic pipeline}

\classification{95.75.Wx -- 97.10.Sj}
\keywords      {Stellar pulsations -- Solar-like oscillations -- Data analysis }

\author{S. Mathur, }{
  address={Indian Institute of Astrophysics, Koramangala, Bangalore-560034, INDIA}
}

\author{R.~A. Garc\'ia}{
  address={ Laboratoire AIM, CEA/DSM -- CNRS - Universit\'e Paris Diderot -- IRFU/SAp, 91191 Gif-sur-Yvette, France}
}

\author{C. R\'egulo}{
  address={Instituto de Astrof\'isica de Canarias, 38205, La Laguna, Tenerife, Spain}
  ,altaddress={Universidad de La Laguna, 38206 La Laguna, Tenerife, Spain
} 
}

\author{J. Ballot}{
  address={ Laboratoire d'Astrophysique de Toulouse-Tarbes, Universit\'e de Toulouse, CNRS, F-31400, Toulouse, France}
}

\author{D. Salabert}{
  address={Instituto de Astrof\'isica de Canarias, 38205, La Laguna, Tenerife, Spain}
}

\author{W.~J. Chaplin}{
  address={School of Physics and Astronomy, University of Birmingham, Edgbaston, Birmingham B15 2TT, UK
}
}

\begin{abstract}
Kepler mission will provide a huge amount of asteroseismic data during the next few years, among which hundreds of solar-like stars will be targeted. The amount of stars and their observation length represent a step forward in the comprehension of the stellar evolution that has already been initiated by CoRoT and MOST missions. Up to now, the slow cadence of observed targets allowed an individual and personalized analysis of each star. During the survey phase of Kepler, this will be impossible. This is the reason why, within the AsteroFLAG team, we have been developing automatic pipelines for the Kepler solar-like oscillation stars. Our code starts by finding the frequency-range where p-mode power is present and, after fitting the background, it looks for the mode amplitudes as well as the central frequency of the p-mode hump. A good estimation of the large separation can thus be inferred in this region. If the signal to noise is high enough, the code obtains the characteristics of the p modes by doing a global fitting on the power spectrum. Here, we will first describe a few features of this pipeline and its application to AsteroFLAG synthetic data to check the validity of the code. 
\end{abstract}

\maketitle


\section{Introduction}

Kepler mission, which is dedicated to the search for exoplanets as well to asteroseismic analysis was launched in March 2009. During the survey phase, a few hundreds of solar-like oscillation stars will be observed for one month. To analyse this huge number of stars very quickly, several pipelines have been developed. We present here our pipeline applied to stars simulated by the AsteroFLAG group. It has also been tested on a few solar-like oscillation stars already observed with CoRoT, such as HD49933 \citep{2008A&A...488..705A}, HD181906 \citep{2009arXiv0907.0608G}, and HD181420 \citep{Barba09}.

\section{Pipeline description}


The pipeline is constituted of five main packages:

\begin{enumerate}
\item
  
   P-Mode Range Search: we look for the frequency range of the p-mode excess power as well as the large-frequency separation estimation:
   \begin{itemize}
 \item first, we estimate the large separation, $\Delta \nu$, by calculating the Power Spectrum (PS) of PS from 200 to 6000 $\mu$Hz (called PS2 hereafter, see an example on Fig. 3).
\item then, we take sub-regions of 600 $\mu$Hz in the PS, compute the PS2 and look for the relative power of the highest peak close to the value of $\Delta \nu$ found. By calculating the threshold corresponding to the probability of having a peak at this position, we obtain the range of p-mode excess power ($f_{min}$ and $f_{max}$) with a given confidence level (varying from 70 to 95\% \citep{ChaEls2002})
   \end{itemize}
  \item Background Fitting: we fit the background with 3 components, 2 harvey laws \citep{1985ESASP.235..199H} and a flat photon noise. The technique used is the maximum likelihood minimization.
  \item Bolometric amplitude of the $\ell$ = 0 mode: we look for the amplitude envelope of the PS
  \begin{itemize}
 \item we correct the PSD from background and we smooth it over a window of 1 or 2x$\Delta \nu$; 
\item we fit the result with a Gaussian to get $P_{max}$ and $\nu_{max}$;
\item we convert the power to the amplitude per radial mode $A_{max}$ \citep{2008ApJ...682.1370K} and to bolometric amplitude \citep{2009A&A...495..979M}.   
	\end{itemize}
  
  \item First determination of the Mass and Radius of the star:  we use the inversed scaling laws \citep{1995A&A...293...87K, 2007A&A...463..297S}
  \item Global fitting: we determine the table of guesses of modes frequencies from $\nu_{max}$ and $\Delta \nu$. We use a maximum likelihood minimization fitting of all the modes in a given frequency range including also the background. The code allows several possibilities like including Bayesian constraints, different mode identification and, $\ell$ = 3 modes.
   
\end{enumerate}

The pipeline also contains three more packages:
\begin{enumerate}
\item
Curvelets: Enhance the ridges of the p-mode signal in the echelle diagram \citep{LamPir2006}
\item
PSSPS (Power Spectrum of a Short part of the Power Spectrum), which is also part of package \#1: 
\begin{itemize}
\item
we estimate $\Delta \nu$ from the standard stellar parameters
\item
we search for the spacing in the PS of a short slice of the PS where the p-mode excess power appears \citep{2002A&A...396..745R}
\end{itemize}
\item Determination of the rotation period (with several techniques among which the wavelets)
\end{enumerate}


\section{Results obtained with the AsteroFLAG hare and hounds}

For 3 different stars generated by the asteroFLAG team, we have applied our pipeline (from packages \#1 to 4): Pancho, Boris and Arthur \citep{2008AN....329..549C}. Each star has been simulated with 3 different magnitudes ($m_v$) (or Signal to Noise Ratio, SNR): 9, 10 and 11. They are all dwarves (V). Pancho has a $T_{eff}$=6383 K, for Boris, $T_{eff}$=5780 K and for Arthur, $T_{eff}$=6420 K.
Each time-series is 3 years long. In our analysis, we used sub-series of 31 days as KEPLER will observe this length of data during the survey phase before focusing on a few star for several years.


Fig. 1 shows the results for the estimation of $\Delta \nu$ with package \# 1 for magnitudes of 9 and 11. 
We can see that in most of the cases, we manage to find a good estimation of $\Delta \nu$. For $m_v$=9, in 97\% of the cases, we obtained a good value with uncertainties lower than 3 $\mu$Hz. For $m_v$=10, a good estimation is obtained for 65\% of the cases and 36\% for $m_v$=11. Once we have an estimation on the large separation, we calculate the frequency-range of the p-mode excess power. Table 1 gives a statistical overview of the success ratio concerning the detection of the frequency range where most of the p-mode bump is found. For a bright star, almost 100\% of the power excess is correctly found. But above a SNR of 11, results start to become less and less reliable.

\begin{figure}[h!]

\includegraphics[height=.2\textheight , width=0.4\textwidth]{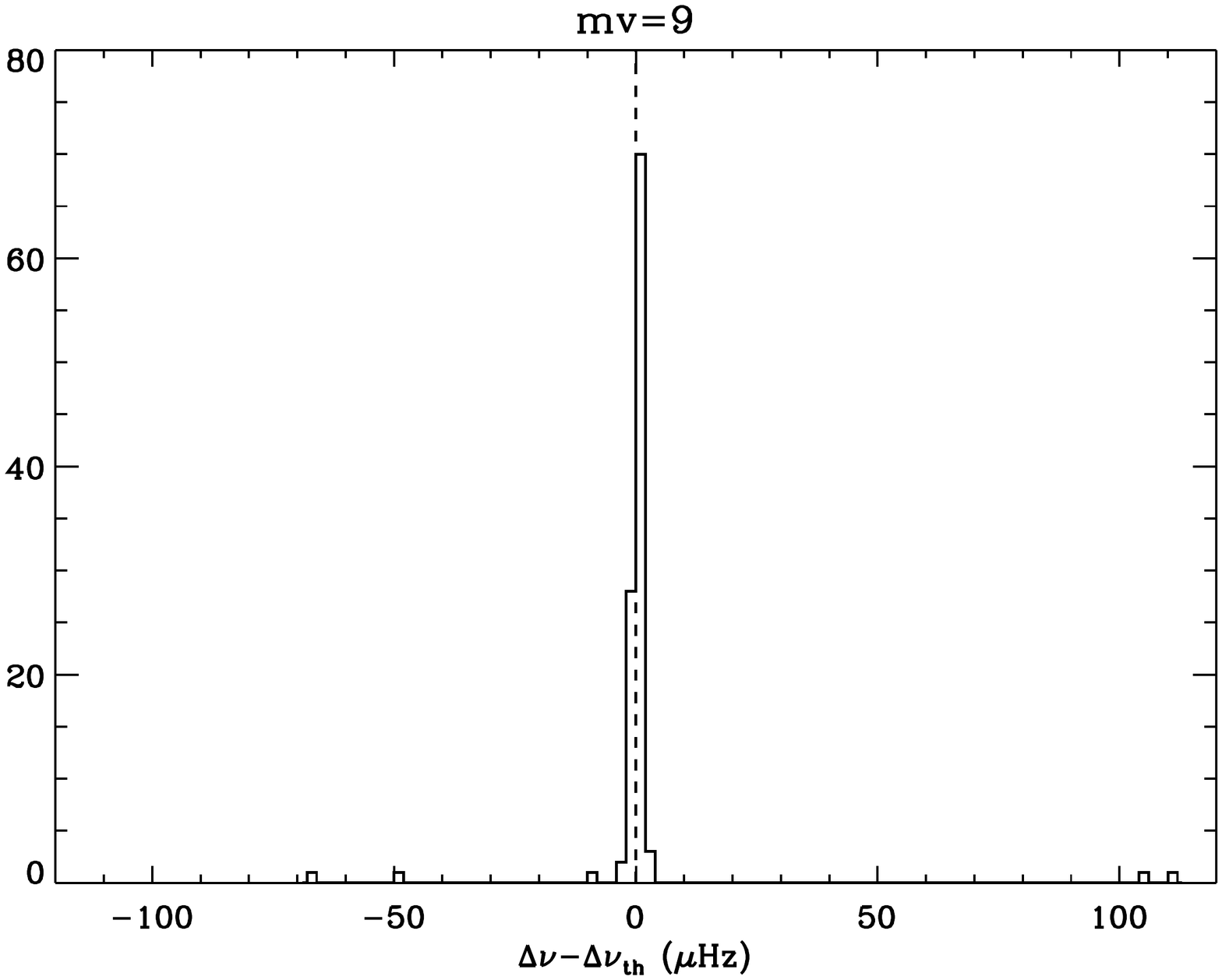}

\end{figure}

\begin{figure}[h]

\includegraphics[height=.2\textheight , width=0.4\textwidth]{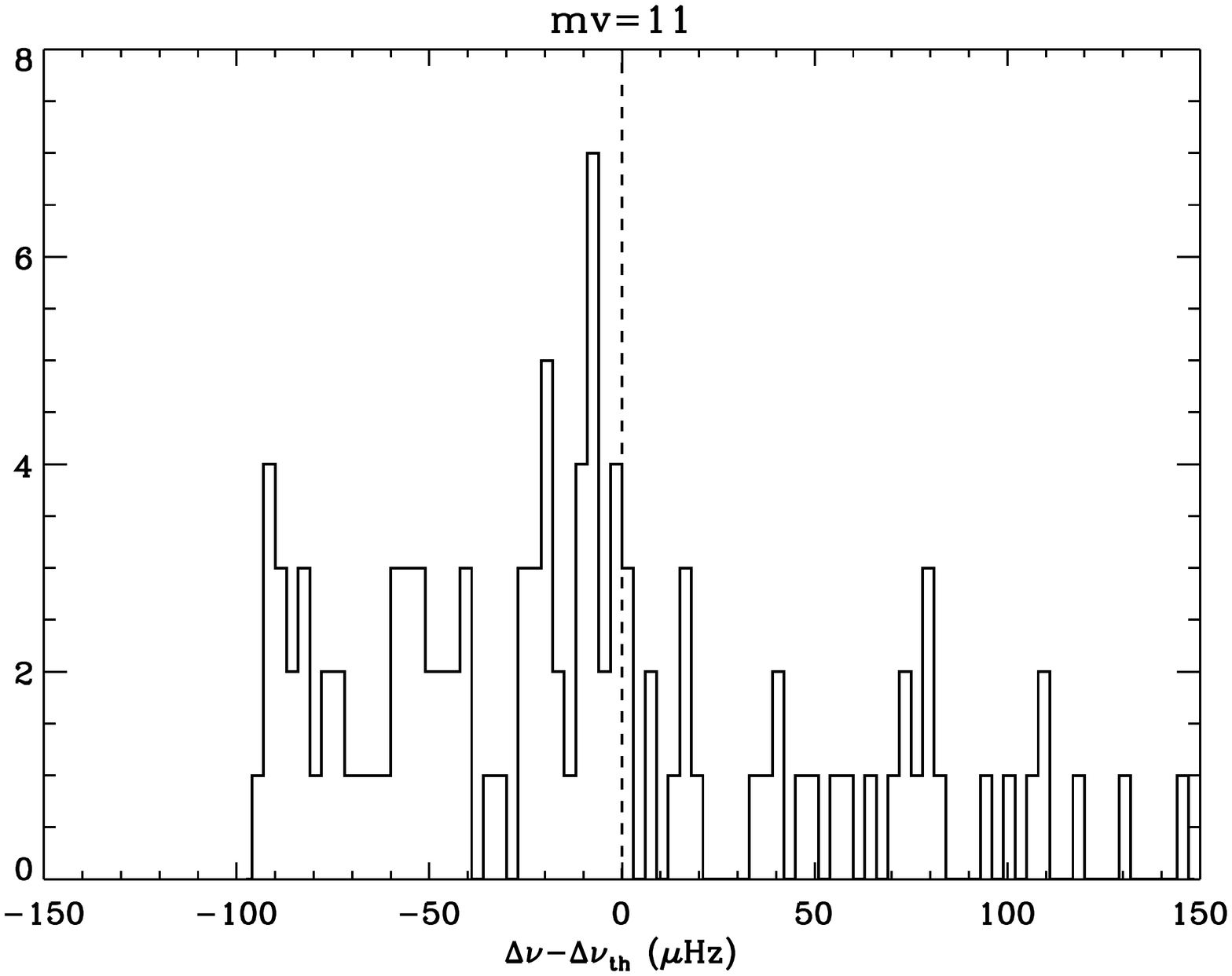}
  \caption{Histograms of the results of the pipeline on the estimation of $\Delta \nu$. For each $m_v$, we have used 3 stars.}
\end{figure}

\begin{table}[h!]
\begin{tabular}{lrrr}
\hline 
  \tablehead{1}{r}{b}{Star/SNR}
  & \tablehead{1}{r}{b}{9}
  & \tablehead{1}{r}{b}{10}
  & \tablehead{1}{r}{b}{11}   \\
\hline
Pancho & 100 & 64 & 28   \\
Boris & 100 & 81 & 61  \\
Arthur & 97 & 61 & 30  \\
All & 99 & 65 & 40  \\
\hline
\end{tabular}
\caption{Percentage of good results for the frequency-range of the p-mode excess power}
\label{tab:a}
\end{table}

Finally, after removing the background fit for each star, we calculate the maximum Amplitude of l=0 mode. Here, we took only the stars where the frequency range was correctly found. The results for $m_v$ =9 and 11 are shown on the histograms of Fig. 2. The amplitude is also well retrieved for most of the stars with SNR of 9 and 10. The error bars associated vary from 0.8 to 2ppm for the 3 SNR. Fig. 3 and 4 show the results obtained for 31 days of Pancho and $m_v$ = 9.


\section{Conclusions}

The pipeline applied to the asteroFLAG hare and hounds datasets gives good results concerning stars with SN of 9 and 10. Within the error bars we retrieve the values of the main global parameters of the modes. 
This pipeline for the packages 1 to 4 is now working in an automated way and was tested on CoRoT targets as well as on simulated stars from the Kepler-asteroFLAG simulator.


\begin{theacknowledgments}
This work has been partially supported by the CNES/GOLF grant as well as ISSI. CoRoT (Convection, Rotation and planetary Transits) is a mini-satellite developed by the French Space agency CNES in collaboration with the Science Programs of ESA, Austria, Belgium, Brazil, Germany and Spain \citep{2006cosp...36.3749B}.
\end{theacknowledgments}

\begin{figure}[h!]

\includegraphics[height=.2\textheight , width=0.4\textwidth]{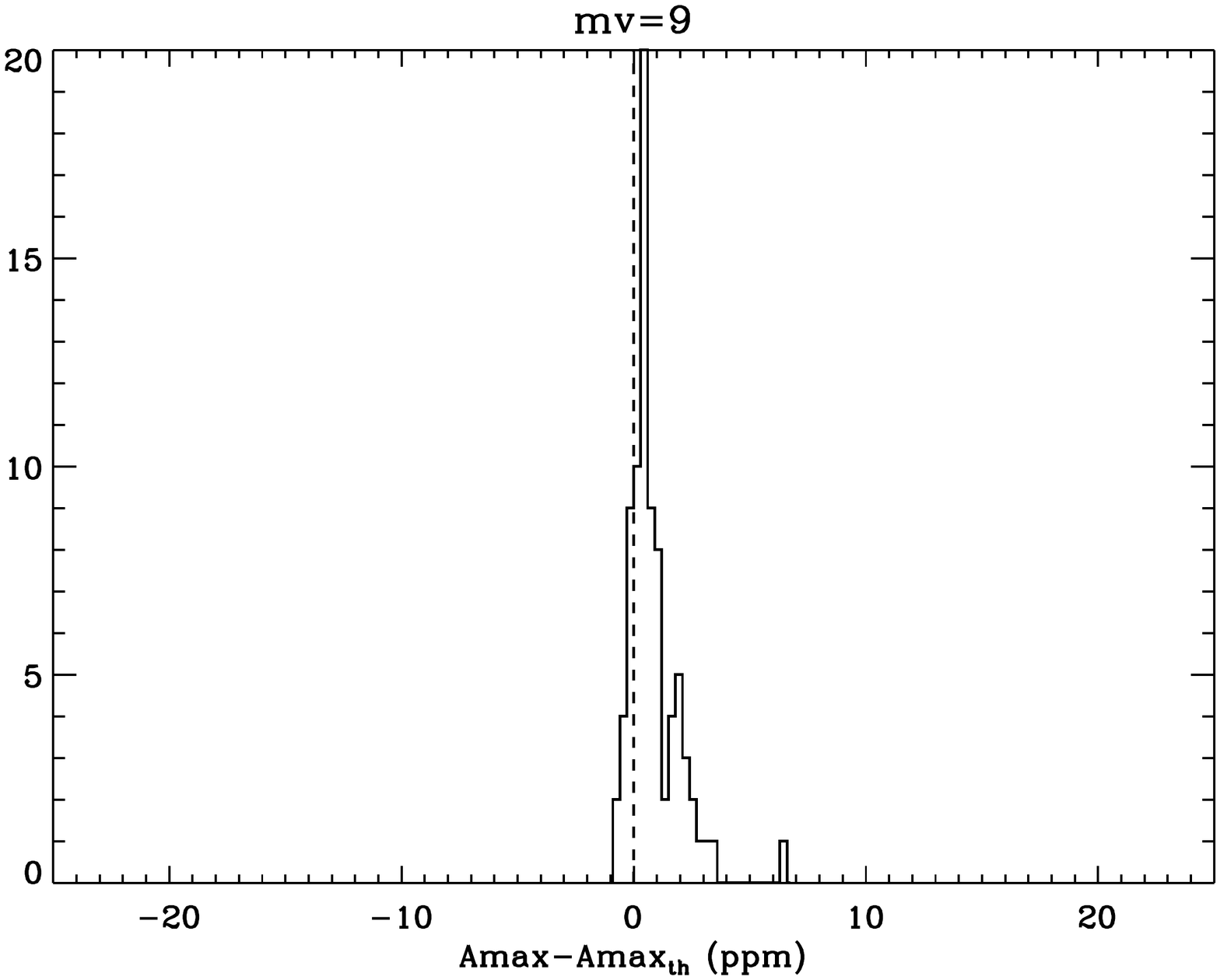}\\

\end{figure}

\begin{figure}[h!]

\includegraphics[height=.2\textheight, width=0.4\textwidth]{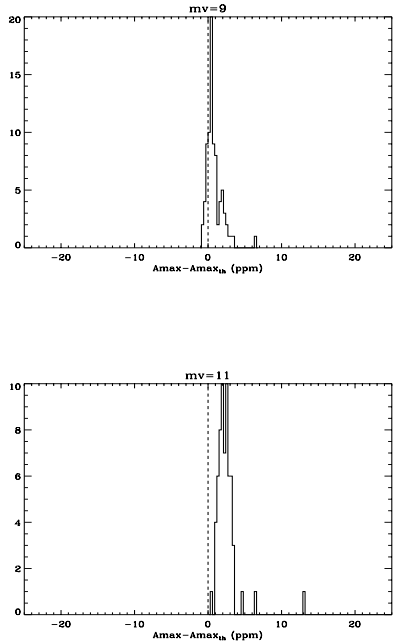}
  \caption{Histograms of the results of the pipeline on the maximum amplitude per radial mode. Same as figure 1.}
\end{figure}

\begin{figure}[h]

\includegraphics[width=.35\textwidth]{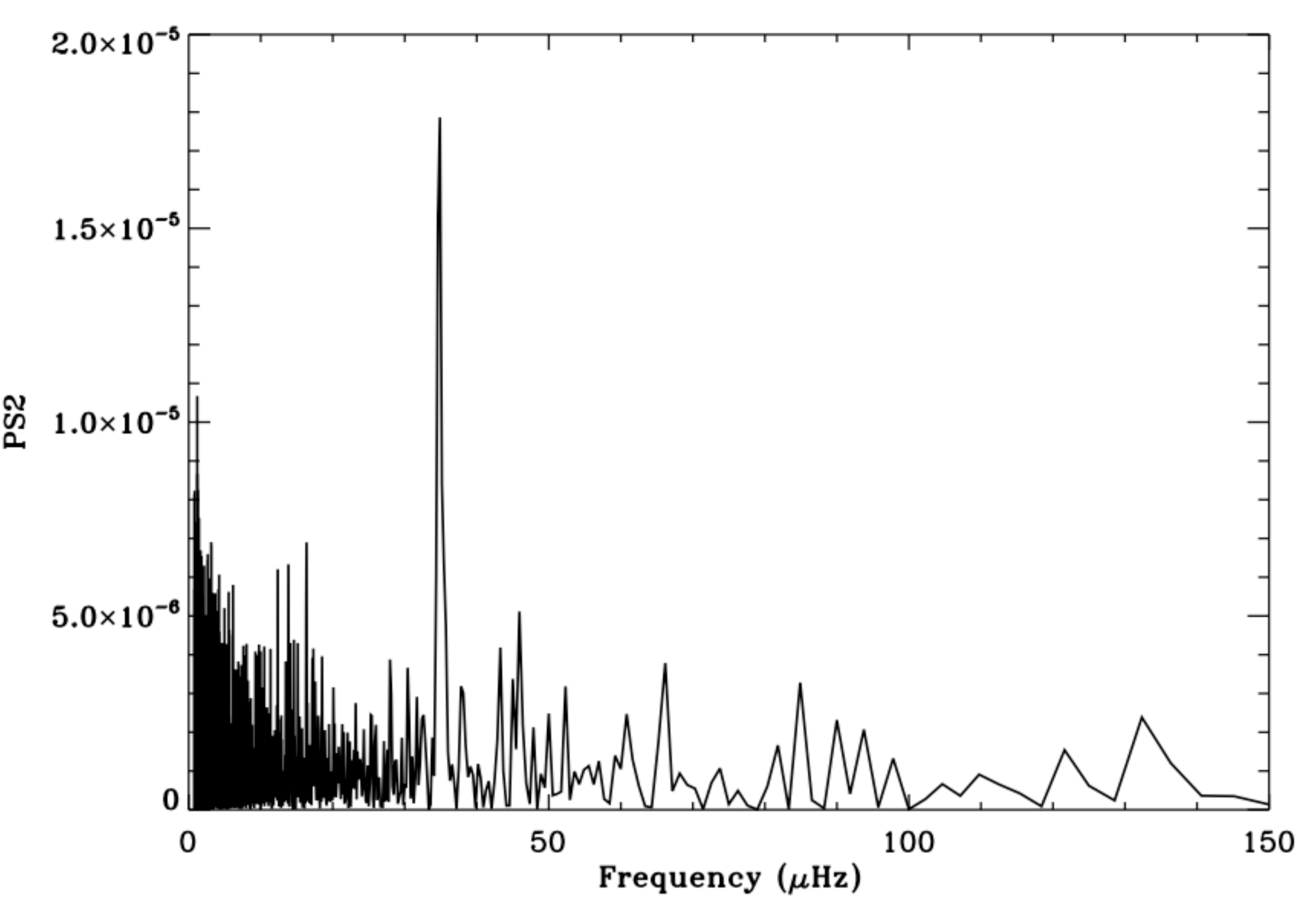}\\

  \caption{Example of the PS2 for 31 days of Pancho.}
\end{figure}

\begin{figure}[h]

\includegraphics[width=.35\textwidth]{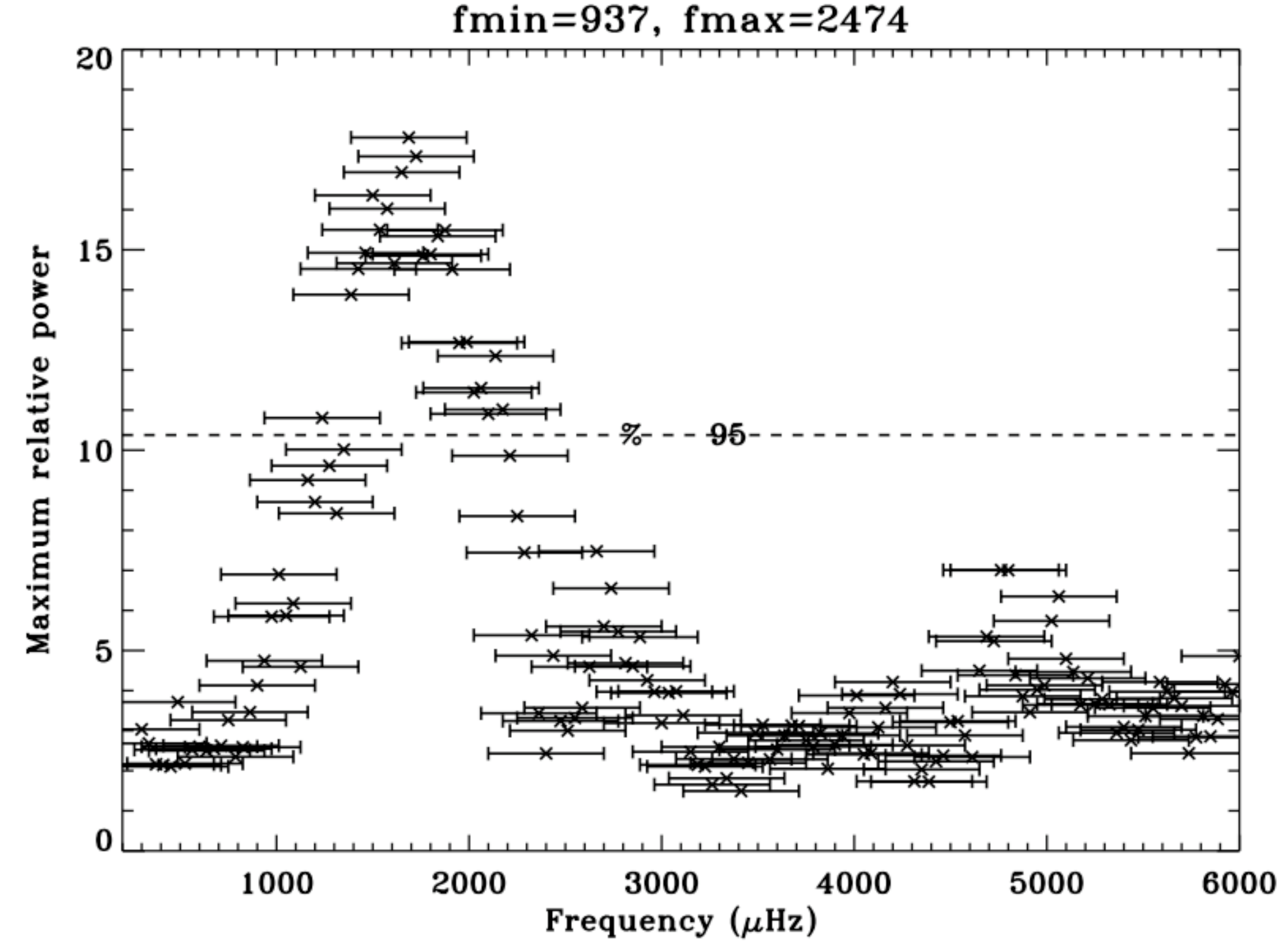}
\end{figure}

\begin{figure}[h]

\includegraphics[height=.1\textheight]{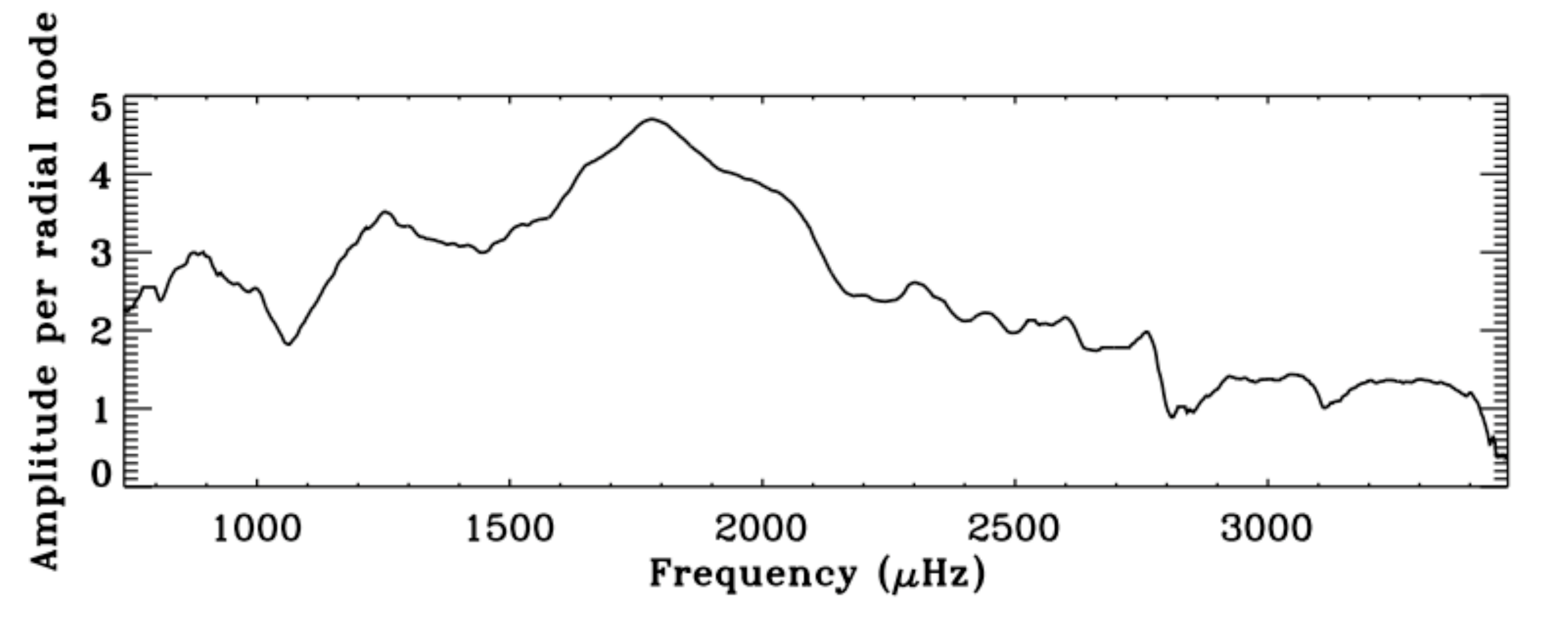}
  \caption{Example of Pancho: results of package \#1 (top) and 4 (bottom).}
\end{figure}



\bibliographystyle{aipproc}   

\bibliography{/Users/Savita/Documents/BIBLIO_sav}

\IfFileExists{\jobname.bbl}{}
 {\typeout{}
  \typeout{******************************************}
  \typeout{** Please run "bibtex \jobname" to optain}
  \typeout{** the bibliography and then re-run LaTeX}
  \typeout{** twice to fix the references!}
  \typeout{******************************************}
  \typeout{}
 }

\end{document}